\documentclass[sigconf]{acmart}
\usepackage[utf8]{inputenc}
\usepackage{graphicx}
\usepackage{cancel}
\usepackage{caption}
\usepackage{subfig}
\usepackage{amsmath}
\usepackage{amsthm}
\usepackage{verbatim}
\usepackage{makecell}
\usepackage{enumitem}
\usepackage[linesnumbered,ruled,vlined]{algorithm2e}

\newcommand{\ie}{\textit{i}.\textit{e}., }
\newcommand{\eg}{\textit{e}.\textit{g}., }
\newcommand{\aka}{\textit{a}.\textit{k}.\textit{a}., }
\newcommand{\etc}{\textit{etc}. }
\newcommand{\resp}{\textit{resp}. }

\newcounter{myexample}

\begin{document}

\title{ZodiacEdge: a Datalog Engine With Incremental Rule Set Maintenance}

\author{Weiqin Xu}
\affiliation{%
  \institution{Engie Lab, CRIGEN}
  \city{Stains}
  \country{France}}
\email{weiqin.xu.paris@gmail.com}

\author{Olivier Curé}
\affiliation{%
  \institution{LIGM Univ Gustave Eiffel, CNRS, F-77454}
  \city{Marne la Vallée}
  \country{France}}
\email{olivier.cure@univ-eiffel.fr}

\begin{abstract}
In this paper, we tackle the incremental maintenance of Datalog inference materialisation when the rule set can be updated. This is particularly relevant in the context of the Internet of Things and Edge computing where smart devices may need to reason over newly acquired knowledge represented as Datalog rules. Our solution is based on an adaptation of a stratification strategy applied to a dependency hypergraph whose nodes correspond to rule sets in a Datalog program. Our implementation supports recursive rules containing both negation and aggregation. We demonstrate the effectiveness of our system on real and synthetic data.
\end{abstract}

\maketitle

\section{Introduction}
In the field of database management systems, Datalog\cite{DBLP:books/aw/AbiteboulHV95} pertains to a declarative query language that is gaining traction. This language has found acceptance among both commercial and open-source systems, as demonstrated by its adoption in Datomic\footnote{https://www.datomic.com/}, 
Oracle's RDF store \cite{oracleRDF}, RDFox \cite{rdfox}, Yedalog \cite{yedalog}, RelationalAI\footnote{https://relational.ai/}, BigSR \cite{bigSR} and BigDatalog \cite{bigDatalog} to name a few. A key feature of Datalog is that it supports the derivation of implicit consequences from explicit data and knowledge. Although these inferences can be computed at query execution time, many systems favour their materialisation to speed up query response. Nevertheless, the efficiency of materialisation has some drawbacks when it comes to modifications to the underlying information.

These limitations have motivated research that has led to the design of incremental reasoning approaches.
Briefly, the aim of these algorithms is to modify part of the materialisation when updates occur, without recalculating inferences from the original data set.
Several research papers, \eg \cite{DBLP:conf/sigmod/GuptaMS93}, \cite{Motik2015IncrementalUO}, have proposed incremental solutions to efficiently update a materialisation for data-level changes, \ie when new data is inserted or when some data is removed from the fact set.
To the best of our knowledge, no incremental maintenance solution has been proposed when updates concern the rule set, \ie when new rules are inserted into or deleted from the Datalog program. 

One type of use case where these updates are relevant involves the Internet of Things (IoT) and Edge computing\cite{edgeComp}, a processing paradigm that brings the storage, management, and processing of data closer to where it needs to be done. In fact, we have recently encountered industry scenarios (detailed in Section \ref{motivation}) in which the rule set of an autonomous device evolves due to the reception of additional rules, from either servers or IoT devices, thus enriching the devices' reasoning capabilities . 
In this type of context, rules can be added and/or removed from a Datalog program and both operations involve the insertion or removal of facts in the materialisation. 

In our Datalog engine solution, we support recursion, disjunction, negation, and aggregation operations. 
The expressiveness of this language makes it particularly difficult for a naive approach to efficiently maintain correct data states.
In this work, we present an incremental approach in which the rules of the Datalog program can be updated. Note that this implies that our engine also supports incremental dataset maintenance.

The key element of our approach is a dependency hypergraph whose hypernodes consist of Datalog rules that belong to the same stratum according to the standard Datalog stratification approach \cite{przymusinski1989declarative}. Dedicated evaluation algorithms, based on the semi-naive approach \cite{DBLP:books/aw/AbiteboulHV95}, have been designed and implemented in our ZodiacEdge prototype.
This system currently accepts Datalog programs whose predicates have a maximum arity of two. This is motivated by our need to support several projects of our industrial partner, ENGIE, which makes extensive use of RDF (Resource Description Framework) data. However, our system can easily be adapted to Datalog programs with unbounded predicate arity.

This paper is organised as follows. In the next section, we present some background knowledge on Datalog and RDF. In Section \ref{motivation}, we motivate our approach and implementation in the context of an industrial Edge computing scenario. Section \ref{zodiacEdge} presents the main components and algorithms of our prototype. Section \ref{relatedWork} presents some related work in the field of Datalog incremental maintenance. In Section \ref{evaluation}, a detailed experiment evaluates our system on both a real world and on several synthetic use cases. It highlights that our approach can be up to three orders of magnitude more efficient than a complete recomputation of a materialisation. We conclude this paper and present some future work in Section \ref{conclusion}.

\section{Preliminaries}
\subsection{Datalog}
We now present the language of Datalog with negation where a rule is an expression of the following form:
\begin{equation}
H \longleftarrow B_1 \wedge B_2 ... \wedge B_{k} \wedge not B_{k+1} ... \wedge not B_n 
\label{datalog_syntax}
\end{equation}
    
The elements of (1), \ie  $H, B_1,..,B_k,..,B_n$,  are atoms corresponding to formulas of the form $P(t_1,..,t_n)$ where P is a predicate and each $t_i$ is a term which can be a variable or a constant. An atom is a fact when all the terms are constants. The atom H is the head of the rule and the conjunction of $B_i$ atoms is the body of the rule. The sets $\{B_1, \ldots, B_k\}$ and $\{B_{k+1}, \ldots, B_n\}$ represents respectively the sets of positive and negative body atoms. It is required for each rule $r$ to be safe, meaning that every variable appearing in $r$ must also appear in at least one positive body atom. A rule is interpreted as a logical implication where, if the body of the rule is true, then the head is also true. The extensional database, or EDB \cite{DBLP:books/aw/AbiteboulHV95}, is made up of all the tuples corresponding to explicit facts in the program, using so-called EDB predicates. The intensional database, also known as the IDB \cite{DBLP:books/aw/AbiteboulHV95}, consists of tuples that are inferred by the program rules and associated with IDB predicates. These IDB predicates are defined in the head of a rule. A program is a finite set of rules.

The evaluation of a Datalog program corresponds to the repeated application of its rules on data tuples until this process reaches a fixed point. Several evaluation methods have been proposed in \cite{DBLP:books/aw/AbiteboulHV95}. In this work, we adopt the semi-naive method which aims to minimize the number of redundant derivations that are computed. This is achieved by ensuring that each deduction in a given iteration implies at least one fact derived in the previous iteration.

The most straightforward semantics for Datalog with negation is stratified negation. It partitions a Datalog program into a sequence of strata of semipositive Datalog programs, \ie composed of rules where negation appears only in rule bodies and solely on EDB predicates. During the evaluation, the strata are evaluated in topological order and the resulting IDB of a stratum s-1 is considered to be the EDB of the next stratum s.



We now formally define a Rule Dependency Graph, henceforth $RDG$, \ie a tool for optimising and efficiently executing Datalog programs. A $RDG$ corresponds to a tuple $(V,E_p,E_n)$ where V is a finite set of vertices corresponding to Datalog rules, $E_p$ and $E_n$ are finite sets of edges corresponding to dependency relationships between these rules. Given a set of rules $R$ where all rules are following the form of (\ref{datalog_syntax}), a $RDG$ is constructed as follows:
\begin{itemize}
    \item RDG.V = R
    \item Considering two rules $r_i,r_j \in R$, if a predicate $p$ appears in the head of a rule $r_i$ and also appears as a positive predicate in a rule $r_j$'s body then $(r_i, r_j) \in RDG.E_p$, \ie is a positive dependency.
    \item Considering two rules $r_i,r_j \in R$, if a predicate $p$ appears in the head of a rule $r_i$ and also appears as a negative predicate in a rule $r_j$'s negative body then $(r_i, r_j) \in RDG.E_n$, \ie is a negative dependency.
\end{itemize}

\begin{figure}[ht]
\centering
\includegraphics[scale=0.46]
{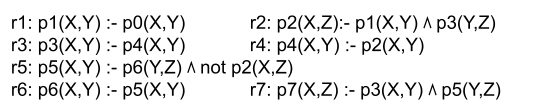}
\caption{Running rule set example}
\label{fig:original_rule_base}
\end{figure}

\begin{figure}[ht]
\centering
\includegraphics[scale=0.48]{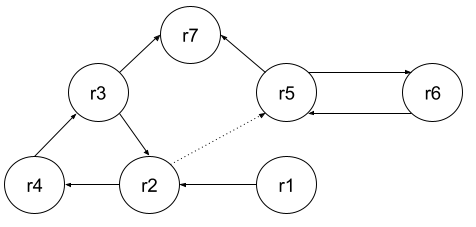}
\caption{Rule dependency graph corresponding to Fig. \ref{fig:original_rule_base}}
\label{fig:original_rdg}
\end{figure}

Given the set of rules of Fig. \ref{fig:original_rule_base}, we can generate the $RDG$ presented in Fig. \ref{fig:original_rdg}, where solid and dotted arrows respectively correspond to positive and negative dependencies. Fig. \ref{fig:original_rdg} can be separated into two graphs: one with only positive dependencies made up of the $RDG.E_p$ and another graph with only negative dependencies consisting of $RDG.E_n$.

Traditional Datalog reasoning uses stratification to help determine a topological order of rules based on the appearance of negation. 
Given a set of Datalog rules $R$, a stratification is a function $f: R \rightarrow \mathbb{N}$ where for $r \in R$, $f(r)=n$ states that rule $r$ is in the stratum number $n$. We also define the set of rules of a given stratum as $\mathcal{S}_n = \{r|f(r) = n\}$. Given the $RDG$ of $R$, the function $f$ must satisfy two constraints:
\begin{itemize}
    \item $\forall r_i,r_j \in R$, if $(r_i,r_j) \in RDG.E_p$, then $f(r_i) \le f(r_j)$.
    \item $\forall r_i,r_j \in R$, if $(r_i,r_j) \in RDG.E_n$, then $f(r_i) < f(r_j)$.
\end{itemize}

In order to guarantee reasoning's correctness, this strategy computes the result stratum by stratum following a bottom-up approach.

\begin{figure*}[ht]
\centering
\includegraphics[scale=0.42]{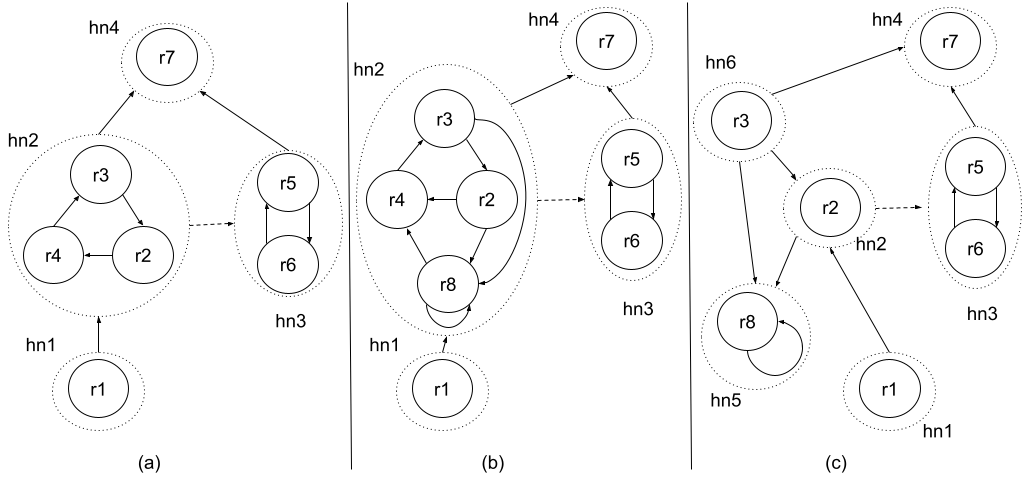}
\caption{The Hyper rules dependency graph corresponding to (a): Fig. \ref{fig:original_rdg}, (b): the addition of rule r8, and (c): the deletion of rule r4 (c)}
\label{fig:hrdg}
\end{figure*}

\subsection{RDF}
The Resource Description Framework (RDF) \cite{rdf11} corresponds to the data model that is generally supporting the design of a Knowledge Graph (KG)\cite{DBLP:journals/csur/HoganBCdMGKGNNN21}. It is taking the form of a labeled, directed multi-graph. Assuming disjoint infinite sets I (IRI for Internationalized Resource Identifiers), B (blank nodes), and L (literals), a triple (S,P,O) $\in$ (I $\cup$ B) x I x (I $\cup$ B $\cup$ L) is called an RDF triple where S, P, and O respectively denote the subject, property, and object of that triple. 

RDF is the cornerstone of the Semantic Web and supports the creation of Knowledge Bases (KBs).
A KB consists of an ontology, \aka terminological box (TBox), and a fact base, \aka assertional box (ABox). 
The TBox supports reasoning mechanisms. Several ontology languages have been designed by the W3C. They differ by their expressive power, \ie the inferences they can compute. 
As far as reasoning with KGs is concerned, Datalog is commonly used to answer queries about ontologies. For instance, one can use Datalog to answer queries over the RL (Rule language) profile of OWL (Web Ontology Language)\cite{owlRL} ontologies which may also include SWRL (Semantic Web Rule Language) rules \cite{swrl}.


 

\section{Motivation}
\label{motivation}
We present an industrial use case related to the business of ENGIE, a multinational company operating in areas such as energy transition, generation and distribution. We consider a smart grid architecture where servers in the Cloud store KGs while Edge devices store their own RDF graphs containing recent data from their sensors. Concerning the latter, the local data storage is essential to maintain data privacy. 

The main motivation for adopting the RDF data model is the existence of ontologies such as Sensor, Observation, Sample, Actuator (SOSA\footnote{http://www.w3.org/TR/ns/sosa}), Quantities, Units, Dimensions, and Types (QUDT\footnote{http://qudt.org/schema/qudt}) or Smart Applicances Reference (SAREF\footnote{https://ontology.tno.nl/saref.ttl}), which facilitate the design of semantic-based applications.

In this context of Edge computing, devices at the edge of the network enable low-latency decision making through inference-based processing.
As a result, the devices are equipped with a Datalog engine. However, Edge devices are often resource-limited, so it is usually impossible to load all the Datalog rules from the Cloud to the Edge device.
In addition, it may not be necessary to load all these rules onto an Edge device, as many rules are only used in certain cases, such as rules for analysing certain anomalies or emergency situations. Consequently, a Datalog engine for Edge computing must have the ability to update its rule set, preferably incrementally. When it encounters certain events, it requests knowledge (Datalog rules) from the Cloud or nearby devices to obtain the necessary rules. Once the problems have been resolved, the local instance of ZodiacEdge can delete the unnecessary rules and retain only a minimal set of rules to manage its day-to-day work.

We have encountered this type of situation in wind farms comprising around thirty turbines producing energy. Some components of these wind turbines collect more than a hundred types of measurements, \eg average wind speed, mechanical torque, which are used to prevent anomalies and predict energy consumption. In some situations, an Edge device may require additional rules to confirm a potential anomaly or provide explanations. 
In a typical situation, a wind turbine $WT_1$ belonging to a certain wind farm obtains an unusual reading from its ambient temperature sensor. Such a problem may have different causes, 1) the sensor may be damaged or broken, 2) the ambient temperature may have changed. To verify the root cause of such an abnormal measurement, the wind turbine can simply compare its ambient temperature measurement with that of other wind turbines belonging to the same wind farm. To do this, $WT_1$ needs to know who its wind turbine neighbours are (this information is provided by servers in the Cloud) and asks them for their ambient temperature. Thanks to additional rules sent by a server in the cloud, $WT_1$ is able to calculate the median ambient temperature of its neighbors and compare it with its own value. Once the cause is found, it can require the corresponding service to handle the problem and return to its original rule set. 

\section{ZodiacEdge system}
\label{zodiacEdge}
In this section, we present the most relevant components of ZodiacEdge with a focus on the incremental processing of rule sets. Specifically, we present our storage strategy, the dependency hypergraph, and the inference processing. 

\subsection{Storage solutions}
In a Datalog engine, data storage can play a crucial role in overall system performance due to the large number of input/output operations. We have designed ZodiacEdge's storage and indexing solutions in a hierarchical way, so that they match our approach to incremental processing of Datalog rules. This approach consists of so-called DataStores and DataStoreBags, which we present below.
\subsubsection{DataStore}

The basic data structure of ZodiacEdge stores Datalog facts in main memory and is called DataStore (DS). To enable fast search operations, a DS is based on two indexes: PSO and POS, where P, S and O correspond respectively to the elements of an RDF triple, i.e. a predicate, a subject and an object. 

This indexing strategy is motivated by the type of queries generally executed on a Datalog engine, \ie the predicate is often a constant while the subject and the object can be variables. These two indexes are stored as hash maps, with P has the key in both cases and SO (resp. OS) corresponding to the value of the PSO index (resp. POS index), hence providing a valuable trade-off between data updating performance and data retrieving efficiency.

\subsubsection{DataStoreBag}
\label{sec:DSB}
Due to the use of rule stratification, situations frequently arise during the reasoning process where the merging of two DS becomes necessary.
For instance, this can occur when combining the extensional database (EDB) of a hyper-node that depends on the intensional database (IDB) of another hyper-node. In Section \ref{hrdg}, we present the notions of a hyper rules dependency graph and hypernodes.
Such a merge operation can be quite expensive for large EDBs and IDBs and has a negative impact on performance when incremental maintenance is required. 

Moreover, once merged, the facts from EDBs and IDBs can not be distinguished. This means that the only way to identify certain facts that need to be deleted is to recalculate the deductions. This recomputation is highly inefficient and does not match with an incremental approach. Instead, we need a solution that unambiguously identifies proofs of certain rules such that they can be removed and/or replaced when some other rules are impacting them.
To avoid this inefficiency, we use a data structure which we called DataStoreBag (DSB). This data structure stores a set of DSs. As a result, it prevents  unnecessary data merging operations. 

Additionally, in order to accelerate data retrieving phase, we implement a predicate-DataStore (p-DS) index which maps each predicate $p_i$ to DSs containing facts that have $p_i$ as predicate. This design is closely related to the definition of Hyper Rule Dependency Graph to support an efficient rule incremental reasoning process. 

\subsection{Hyper Rules Dependency Graph}
\label{hrdg}
In order to define the order in which rules are computed, ZodiacEdge adapts the notion of RDG to a so-called Hyper Rules Dependency Graph (HRDG). This approach forms the foundation of our incremental maintenance approach since it provides a precise execution plan for inference computation.

Intuitively, instead of using strata as reasoning units, ZodiacEdge uses hyper-nodes in HRDG as reasoning units. 

Similarly to RDG, an HRDG contains positive and negative dependencies, respectively denoted as $HRDG.HE_p$
and $HRDG.HE_n$
The difference between RDG and HRDG is that the vertices in RDG are rules, while the vertices of HRDG, denoted $HRDG.HV$, are subgraphs of rules.  Given a set of rules $R$ and its corresponding rule dependency graph $RDG$, we construct an HDRG as follows:

\begin{itemize}
    \item We define a $HRDG = (HV, HE_p, HE_n)$ where $HV$ represents its hyper-nodes, $HE_p$ represents its positive hyper relations and $HE_n$ represents its negative hyper relations. 
    \item $HV$ consists of all the strongly connected components of $RDG$. 
    \item $\forall hn_i, hn_j \in HV$ and $hn_i \neq hn_j$, if $\exists r \in hn_i, \exists r\prime \in hn_j$ and $(r,r\prime) \in RDG.E_p$, then $(hn_i, hn_j) \in HRDG.HE_p$. 
    \item $\forall hn_i, hn_j \in HV$, if $\exists r \in hn_i, \exists r\prime \in hn_j$ and $(r,r\prime) \in RDG.E_n$, then $(hn_i, hn_j) \in HRDG.HE_n$.
    \item To ensure the program's safety, $HRDG.HE_p \cup HRDG.HE_n$ must be a directled acyclic graph (DAG). More precisely, if ZodiacEdge detects that a newly added rule makes the program unsafe, it will reject the insertion of this rule.
\end{itemize}

Through these hyper-nodes, we can clearly identify the dependencies between the different reasoning units. Thus, when processing each reasoning unit, these dependencies can help to reduce the search scope with respect to stratification. Furthermore, since these hyper-nodes compose a DAG, we can completely remove any hyper-node with its IDB without considering the cyclic dependency.

\begin{figure}[ht]
\centering
\includegraphics[scale=0.31]
{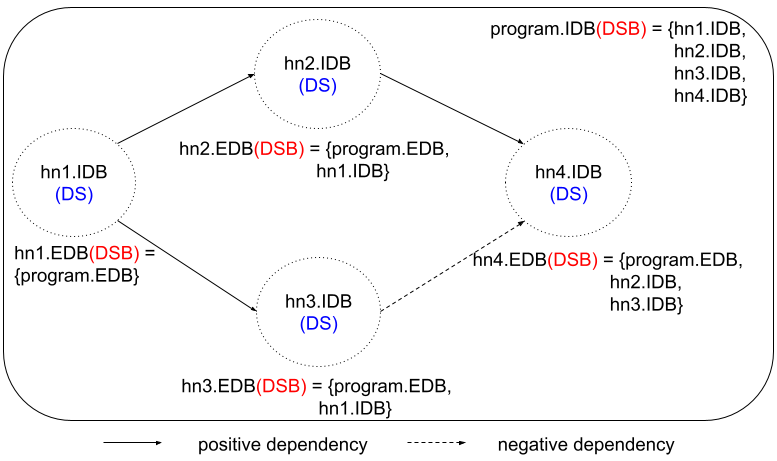}
\caption{DS and DSB organizations for the example of Fig. \ref{fig:original_rule_base}}
\label{fig:ds_dsb_example}
\end{figure}

In Fig. \ref{fig:hrdg}(a), we presented the $HRDG$ of Fig. \ref{fig:original_rdg}'s example. This $HRDG$ contains four hyper-nodes (namely hn1, hn2, hn3 and hn4) where:\\
$HRDG.HV = \{hn1,hn2,hn3,hn4\}$, $HRDG.HE_p = \{(hn1,hn2),(hn2,$\\$hn4),(hn3,hn4)\}$ and $HRDG.HE_n = \{(hn2,hn3)\}$.1

Specifically, the EDB of the reasoner is a DS, while its IDB is a DSB. Moreover, the EDB of a hyper-node is a DSB while its IDB is a DS. Finally, the program.EDB is a DS that stores all the ground truth of the program. Fig. \ref{fig:ds_dsb_example}  gives a concrete example of a Datalog program with 4 hyper-nodes (noted hn1,hn2,hn3,hn4). The EDB of a hyper-node is a DSB that stores a collection of its dependent IDBs with program.EDB. The IDB of a hyper-node is a DS that stores all inferred results produced by the rules of that hyper-node. Finally, program.IDB is a DSB that collects all the IDBs of the hyper-nodes. As mentioned in the section on DSBs, all DSs in a DSB are indexed by predicates in those DSs to speed up query processing.

\subsection{Reasoning process}

During the reasoning process, the execution plan at the hyper-node level is calculated according to the topological order of $HRDG.HE_p \cup HRDG.HE_n$, and the semi-naive algorithm is used to compute the derivations.
For each hyper-node, we initiate its proper EDB and IDB. The EDB is a DSB containing all the ground truth, \ie other nodes' IDBs and program's EDB, required by this hyper-node. 
The IDB is a DS containing all the deduced knowledge of this hyper-node. 

\begin{algorithm}
    \caption{Reasoning process with rules}
    \label{algo:reasoning_no_incre}
    \LinesNumbered
    \KwIn{Rules $R$, Explicit Data Base $program.EDB$}
    \KwOut{Implicit Data Base $program.IDB$}
    \BlankLine
    Construct Rules Dependency Graph $RDG$ from $R$\;
    Compute set $HV$ that contains all the strongly connected components in $RDG$\;
    Construct Hyper Rules Dependency Graph $HRDG$ from $HV$ and $RDG$\;
    Compute topological order $O$ of $HRDG$\;
    $program.IDB = \emptyset$\;
    \For{$hn$ in $O$}{
        $hn.EDB = (\bigcup \{hn^{\prime}.IDB | hn^{\prime} \in HRDG.HV, (hn^{\prime},hn) \in HRDG.HE_p \cup HRDG.HE_n\}) \cup program.EDB$\;
        
        $hn.IDB$ = Apply semi-naive algorithm to rules in $hn$ to reason based on data $hn.EDB$\;
        $program.IDB = program.IDB \cup hn.IDB$
    }
    return $program.IDB$\;
\end{algorithm}
In Algo. \ref{algo:reasoning_no_incre}, lines 1 to 3 contribute to the construction of the program's HRDG. Given this hyper graph, line 4 creates the topological order which is stored in variable $O$. Given an empty IDB (line 5), we begin (line 6) the reasoning process over each hyper-node $hn$ in the order defined in  $O$.
Then, for each hyper-node $hn^{\prime}$ whose IDB contains evidence supporting rules in $hn$, we place a pointer to $hn^{\prime}.IDB$ into $hn.EDB$ (line 7). Note that $hn.EDB$ must contain $program.EDB$ in case the newly added data contains evidence that supports some rules in $hn$. Having defined $hn.EDB$, we apply the semi-naive algorithm recursively on all the rules in $hn$ to compute $hn.IDB$ (line 8), and finally add the DS $hn.IDB$ to the DSB $program.IDB$ (line 9).
Note that in this algorithm, the $\cup$ operation does not systematically merge the data into the  corresponding DSs (\eg $program.EDB$, $hn.IDB$, \etc), but either adds the pointer of a DS to a DSB or merges sets of DSs' pointers in DSBs.
As mentioned above, this improves the efficiency of reasoning, as merging two indexed DSs (especially large ones) can be a costly operation.

\subsection{Rule-incremental Reasoning process}

In a non-incremental approach, when we update, \ie insertion or deletion of rules, the rule set, the strategy consists in recomputing the whole reasoning process to obtain a new IDB suitable for the modified rule set. Obviously, this is 
not an efficient strategy. Instead, we propose an incremental strategy that is much more efficient.

Using the HRDG of a Datalog program, we can identify the hyper-nodes that are directly affected by the updated rules, we call them directly impacted hyper-nodes (DIHNs). As hyper-nodes are the minimum reasoning units in ZodiacEdge, in the current approach, we will redo the reasoning process for DIHNs instead going through an incremental procedure. This due to the fact that the IDBs of DIHNs are often greatly impacted by rule changes where redoing reasoning may be potentially more efficient than doing incremental reasoning. However, as for the successor hyper-nodes of DIHNs that are also impacted by rule changes, we apply the data incremental strategy on them because these hyper-nodes are often slightly impacted by the changes. Next, we consider the identification of DIHNs in the cases of rule insertion and rule deletion.

\subsubsection{Rule-insertion case}
Algo. \ref{algo:rule_insertion_DIHN_identifying} presents a method to identify DIHNs when some rules are inserted into the Datalog program. First, the input $RDG$ integrates the new rules (line 2). Then, the next step is to consider each new rule as a new hyper-node and add all these hyper-nodes to the $HRDG$ (lines 3-7). Then a hypergraph, denoted $HRDG^t$, is computed from the new HRDG (line 8). For each strongly connected component $hn_i$ in $HRDG^t$, if it contains more than one hyper-node of HRDG, we merge all the sub-graphs together with the corresponding EDBs in its multiple inner hyper-nodes into a new sub-graph and give it to $hn_i$ (lines 11). During merging, we add all newly generated hyper-nodes to the set of DIHNs. If a strongly connected component contains only one hyper-node, we simply extract the sub-graph in this inner hyper node to $hn_i$ (lines 14). We add $hn_i$ to DIHN if the inner sub-graph's rules are newly added (line 15). Finally, we obtain a new $HRDG$ from the new rule base. 

Next, we consider a concrete example where the following rule (namely $r8$) is introduced in our running example of Fig. \ref{fig:original_rule_base}. Note that this new rule corresponds to a disjunction since the predicate $p2$ is calculated from $r2$ and $r8$.

\begin{center}
r8:  p2(X, Z) :- p3(X, Y) and p2(Y, Z).    
\end{center}

\begin{algorithm}
    \caption{DIHNs identifying algorithm for rule-insertion cases}
    \label{algo:rule_insertion_DIHN_identifying}
    \LinesNumbered
    \KwIn{New rules $NR$, rules $R$, explicit data base $program.EDB$, rules dependency graph $RDG$, hyper rules dependency graph $HRDG$}
    \KwOut{Directly impacted hyper-nodes $DIHN$}
    \BlankLine

    $DIHN = \emptyset$\;
    Update $RDG$ with $NR$\;
    \For{$r$ in $NR$}{
        \If {$\exists r^{\prime} \in R, (r,r^{\prime}) \in RDG.E_p$}{
            
            Enrich $HRDG.HE_p$ with relation $(r, hn)$ where $r^{\prime} \in hn $\;
            
        }
        \If {$\exists r^{\prime} \in R, (r,r^{\prime}) \in RDG.E_n$}{
            
            Enrich $HRDG.HE_n$ with relation $(r, hn)$ where $r^{\prime} \in hn $\;
            
        }
    }

    Compute the set of strongly connected components $HV$ of $HRDG$ and  store it in $HRDG^t$

    \For{$hn$ in $HRDG^t.HV$}{
        \eIf {$hn$.size $>$ 1}{

            $hn = \bigcup \{v| v \in hn\}$ (Here we extract the subgraphs from all the vertice $v$ ($v \in HRDG.HV$) in the strongly connected component $hn$ and merge these subgraphs into a single hyper-node. During this procedure, we are also going to merge all the $v.EDB$ into $hn.EDB$)\;
            $DIHN$.add($hn$)\;
        }{
            $hn = hn$'s vertice (Here, $hn$ contains only one hyper-node $v$, thus we replace $v$ in $hn$ by the subgraph in $v$) \;
            \If {$\exists r \in hn, r \in NR$ }{
                $DIHN$.add($hn$)\;
                Compute EDB of $hn$\;
            }
        }
    }

    $HRDG = HRDG^t$\;
    return $DIHN$\;
\end{algorithm}

 With Algo. \ref{algo:rule_insertion_DIHN_identifying}, we first compute a $HRDG^{t}$. After merging strongly connected components, the new $HRDG$ of our running example is displayed Fig. \ref{fig:hrdg}(b) where the hyper-node $hn2$ will be registered into $DIHN$.

\subsubsection{Rule-deletion case}
Algo. \ref{algo:rule_deletion_DIHN_identifying} introduces how to identify DIHNs when some rules are removed from the Datalog program. Let denote with $DR$ the set of rules which are being removed from a program. In line 2, we are effectively removing all $DR$ rules from the program RDG. 
In lines 3 to 7, we create a list of hyper-nodes containing the rules to be deleted. Then, for each hyper-node that has been modified, we register their successors into $DIHN$ (line 9). After that, we recompute strongly connected components of the graphs in these hyper-nodes and split or delete them accordingly (line 10-18). Following the insertion example, we will now provide a concrete deletion example:

Considering the deletion of rule $r4$ from our running example, a new $RDG$ is computed by removing rule r4. The algorithm first registers $hn3$ and $hn4$ into $DIHN$, then splits $hn2$ as is shown in Fig. \ref{fig:hrdg}(c). After the split, we will register $hn2$, $hn5$ together with $hn6$ in $DIHN$. Thus, we obtain the following $DIHN$: $\{hn2, hn3, hn4, hn5, hn6\}$.




\begin{algorithm}
    \caption{DIHNs identifying algorithm for rule-deletion cases}
    \label{algo:rule_deletion_DIHN_identifying}
    \LinesNumbered
    \KwIn{Rules to be deleted $DR$, Rules $R$, Explicit Data Base $program.EDB$, Rules Dependency Graph $RDG$, Hyper Rules Dependency Graph $HRDG$}
    \KwOut{Directly influenced hyper-nodes $DIHN$}
    \BlankLine

    $DIHN = \emptyset$\;
    Remove $DR$ from $RDG$ \;
    $HRDG^t = HRDG$\;
    Dirty hyper-nodes $DHN = \emptyset$\;
    \For{$r$ in $DR$}{
        Add $hn$ to $DHN$ where $r \in hn$\;
        Delete $r$ from $hn$\;
    }

    \For{$hn$ in $DIHN$}{
        Add all successors of $hn$ in $HRDG^t$ to $DIHN$\;
        \eIf{$hn = \emptyset$}{
            Delete $hn$ from $HRDG^t$\;
            \If{$hn \in DIHN$}{
                Delete $hn$ from $DIHN$\;
            }
        }{
            Split $hn$ by recalculating strongly connected components using RDG\; 
            Update $HRDG^t$ with split hyper-nodes\;
            Recompute EDB of each split hyper-nodes\;
            Add these split hyper-nodes to $DIHN$\;
        }
    }
    $HRDG = HRDG^t$\;
    return $DIHN$\;

\end{algorithm}



\begin{algorithm}
    \caption{Compute hyper-nodes execution plan from DIHNs}
    \label{algo:incremental_execution_plan}
    \LinesNumbered
    \KwIn{Directly Influenced hyper-nodes $DIHN$, Hyper Rules Dependency Graph $HRDG$}
    \KwOut{hyper-nodes execution plan $P$}
    \BlankLine

    \SetKwFunction{Dfs}{DepthFirstSearch}
    \SetKwProg{Fn}{Function}{:}{}
    \Fn{\Dfs{$n$, $Visited$, $HRDG$, $P$}}{
        $Visited$.add($n$)\;
        \For{$v$ in $\{n^\prime|(n,n^\prime) \in HRDG.HE_p \cup HRDG.HE_n\}$}{
            \If{$v \notin Visited$}{
                \Dfs{$v$, $Visited$, $HRDG$, $P$}\;
            }
        }
        $P$.append($n$)\;
        
    }
    \textbf{End Function}

    $Visited = \emptyset$\;
    Initiate $P$ as a queue\;

    \For{$n$ in $DIHN$}{
            \If{$n \notin Visited$}{
                \Dfs{$n$, $Visited$, $HRDG$, $P$}\;
            }
        }

    Reverse $P$\;
    return $P$\;
\end{algorithm}

\subsubsection{Incremental reasoning}
Based on these DIHNs, by applying Algo. \ref{algo:incremental_execution_plan}, we can find a hyper-node execution plan $P$ containing all impacted (directly and indirectly) hyper-nodes. The general idea is to find a topological order in a graph from the hyper-nodes of the DIHN. This topological order contains only the hyper-nodes that can be reached by the hyper-nodes of the DIHN.

The incremental reasoning process of each hypernode varies depending on its membership in the DIHN.
For each hyper-node in the DIHNs, there are two strategies for obtaining a new IDB:

\begin{itemize}
  \item considering a rule-insertion operation, we will reason with the semi-naive evaluation 
  on the newly added rules within the hyper-node (based on $hn.EDB \cup hn.IDB$), and then apply the data incremental strategy based on the changes in the IDB of the hyper-node until a fixpoint is reached, \ie no new facts can be generated.
  \item in the case of a rule-deletion operation,  we will repeat the reasoning for the whole hyper-node. 
\end{itemize}

Once we have a new IDB, we record the difference between the new IDB and the previous IDB. This difference will trigger a chain reaction for non-DIHNs in the execution plan by applying the data incremental algorithm. 

\begin{figure}[ht]
\centering
\includegraphics[scale=0.32]{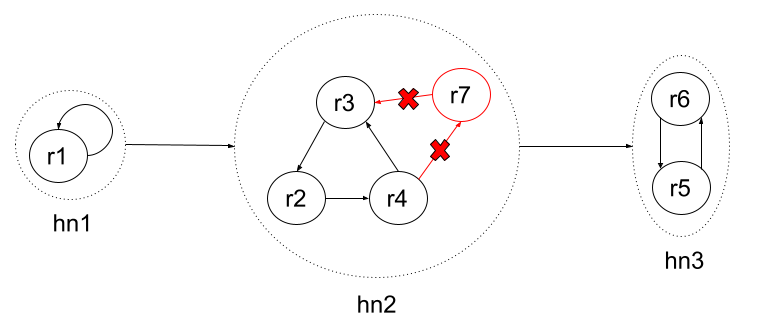}
\caption{A rule-deletion example}
\label{fig:rule_deletion_example}
\end{figure}

As an example, we consider the deletion of rule r7 in Fig. \ref{fig:rule_deletion_example}. This operation does not affect the HRDG's structure. By applying Algo. \ref{algo:rule_deletion_DIHN_identifying}, we find DIHN = \{hn2\}, and the execution plan will consist of hn2 followed by hn3. 
The system will therefore re-process the inferences for hn2. If changes are made to hn2.IDB, the data incremental algorithm is triggered on hn3.

The closer the hyper-nodes are inserted or deleted to the HRDG leaves, the more efficient our rule incremental approach is.
In practical situations we have seen, it often happens that several hyper-nodes are deleted entirely, and during this procedure, the IDBs of these hyper-nodes can be deleted directly without reasoning.

For instance, this is the case when we delete rules r5, r6 and r7 from the original rule base in Fig. \ref{fig:original_rule_base}, where the corresponding hyper-nodes are $hn3$ and $hn4$ in Fig. \ref{fig:hrdg}(a).
Therefore, we will delete $hn3.IDB$ and $hn4.IDB$ from $program.IDB$ ($program.IDB = \{hni.IDB|hni \in HRDG\}$) without reasoning. Our experience in ENGIE's industrial environments has taught us that additional rules, \ie those that are inserted or deleted, usually depend on the rules in an original rule base.

\subsection{Language extensions}
To give our Datalog engine additional functionality for handling more complex cases, we are adding binding, comparison and aggregation operators.
The syntax of the binding operator is as follows:
\begin{equation}
BIND(expression\ AS\ ?v)
\label{binding_operator_syntax}
\end{equation}

The binding operator is considered as an atom in the body of a rule. The expression in (\ref{binding_operator_syntax}) can be composed by basic arithmetic with variables.

The following syntax applies to the comparison operator:
\begin{equation}
COMP(t1,\ comparator,\ t2)
\label{comparing_operator_syntax}
\end{equation}

Here, $t1, t2 \in Variable \cup Constant$ and $comparator \in \{>,>=,=,<=,<\}$. Like the binding operator, the comparison operator is also considered as an atom in the body of a rule.

Finally, the aggregation operator is presented in the form (\ref{aggregation_operator_syntax}) :
\begin{equation}
H\ :-\ AGGREGATE(Bs)\ ON\ ?v1 \\
~WITH\ aggreOp(?v2)\ AS\ ?v3\ .
\label{aggregation_operator_syntax}
\end{equation}


In this syntax, $H$ and $Bs$ respectively correspond to the head and body  of a traditional Datalog rule. $?v1$ and $?v2$ are variables that must appear in $H$, $?v3$ is a variable that usually appears in $H$ and $aggreOp \in \{MAX,MIN,AVG,COUNT,SUM,MED\}$, where the first 5 operations correspond to the standard aggregation operations of SQL. Moreover, the predicate that appears in an aggregation's $H$ can not appear in a second rule's head. This is straightforward because there can not be two different ways of computing the same aggregation. 


\subsection{Rule execution optimization}
Considering the evaluation of a single rule composed of multiple atoms amounts to processing multiple joins between predicates. 
Therefore, just as when executing a database query, the order in which these joins are performed can have a significant impact on system performance.

The execution of ZodiacEdge rules is inspired by the Volcano model \cite{volcano1994}. 
For the join order, ZodiacEdge uses a heuristic strategy and a cost-based strategy to generate an optimised, left-deep join tree. Specifically, ZodiacEdge divides the body of a rule into three parts: positive body, negative body and aggregation. 
A positive body is optimized according to the following heuristic:\\
$P(S,O)>P(?s,O)=P(S,?o)>P(?s,?o)>C(?s)>Bind>Comp$\\
where $P$, $S$ and $O$ correspond to RDF predicate, subject and object constants respectively, \ie IRIs. Moreover, $?s$ and $?o$ are variables.
The $C(?s)$ notation constraints a variable $?s$ to be of a certain type, \ie corresponding to $?s$ \texttt{rdf:type} $C$ in RDF. The $Bind$ operation represents BIND atom (\resp $Comp$ represents COMP atom). 
The order of the predicates in this heuristic is motivated by the respective estimation of their result selectivity. The predicates $Bind$ and $Comp$ are placed at the end of this order because they have a lower impact on the selectivity of the result.


Similarly, we also propose an optimization strategy dedicated to negated predicates:\\
$notP(S,O)>not P(?s,O)=not P(S,?o)>not P(?s,?o)>not C(?s)$

Currently, there can be no more than one aggregation operation in a rule body. This motivates the priority among the three parts is: \\
$PositiveBody > NegativeBody > Aggregation$\\
If two atoms are of the same priority, their ordering will be determined by the cost estimation according to their predicate.

\section{Related work}
\label{relatedWork}

In Delete/Rederive (DRed) \cite{DBLP:conf/sigmod/GuptaMS93}, the plan is to generate two sets of axioms to be included and excluded using a three-part procedure. First, the deletions are determined by identifying the facts that should be removed. This process will lead to an overestimation, because some derived facts can still be inferred from other non-deleted facts. In the next step, the algorithm will identify these facts and remove them from the list of facts to be deleted. Finally, new inferences are made based on the axioms added to the knowledge base in the last step.

In \cite{Motik2015IncrementalUO}, the authors present an approach, denoted B/F, that reduces the overestimation of the first step of DRed. Their approach is based on a combination of backward and forward inference and is available in the RDFox \cite{rdfox} system. ZodiacEdge adapts the B/F algorithm by operating at a finer granularity than  \cite{rdfox},\ie within a given hyper-node. The immediate effect of our approach is that only the relevant set of inferences is modified, without applying the algorithm to the entire IDB.
The ZodiacEdge B/F optimisation is implemented using a dynamic programming (DP) approach. We do not provide details of this algorithm as we consider it to be outside the scope of this paper.
The overall idea of this DP optimized B/F algorithm is similar to F/B/F \cite{DBLP:journals/ai/MotikNPH19}, which tries to combine DRed's overdeletion and backward chaining to limit the effect of overdeletion. However, our DP implementation  algorithm uses memoization within a hyper node, effectively reducing memory usage and eliminating redundant backward checking as much as possible.

A simpler approach to incrementally maintain materialisation consists in associating a counter to inferred facts \cite{10.1145/16856.16861}. The counter of a fact is incremented when new derivation is triggered for that fact, and decremented when that fact no longer holds. The fact is removed when the counter reaches zero. This counting approach has limits with recursive rules over arbitrary data sets. It has been demonstrated in \cite{conf/aaai/HuMH18} that DRed and B/F can successfully be extended with bookkeeping. ZodiaEdge currently does not consider the Counting approach but we believe that it could be beneficial to the overall performance of our system.

In \cite{DBSP}, authors introduce DBSP, an expressive language that supports Incremental View Maintenance(IVM). DBSP supports incrementalization and streaming computation with rich expressiveness that can model various query language classes, \eg SQL and Datalog. However, such incrementalization only focuses on the data-incremental aspect. Comparatively, ZodiacEdge supports both data and rule incremental solutions.%

In \cite{souffle_elastic_incrementalization}, David Zhao et al. present an  elastic incrementalization approach for Datalog. The system has been implemented in the so-called Souffle system. The authors consider that in cases of large updates, a recomputation may be cheaper than an incrementalization. Then, they propose two strategies, namely Bootstrap which recomputes the entire result for high-impact changes and Update which performs an incremental update for low-impact
changes. However, like most related work, this research focuses on the data-incremental cases. 


\section{Experimentation}
\label{evaluation}
\subsection{Evaluation setting and data sets}
Our experimentation has been conducted on a Mac Book Pro with a CPU of 2.6 GHz Intel Core i7 6 cores and 16G RAM. ZodiacEdge is implemented in Python and depends on few libraries that make it easily portable. For instance, we have implemented our own optimized strongly connected components based on Kosaraju's algorithm \cite{SHARIR198167}. The source code of ZodiacEdge can be downloaded from its github page\footnote{https://github.com/xwq610728213/zodiac\_edge}.

We evaluated ZodiacEdge with one real world, namely Rule Set 1 (RS1), and two synthesized, named Rule Set 2 and  3 (resp. RS2 and RS3). RS1 is applied on a data set (DS1) that contains a set of RDF triples in the form of: \\
$<windTurbine_i> \quad <hasNeighbour>\quad <windTurbine_{i+1}>.$ 

The number of wind turbines in this data set ranges from 50 to 400. 
DS1 is typically used to test the performance of ZodiacEdge on symmetric-transitive reasoning, which is the most computationally intensive task among the semantics supported by ZodiacEdge. RS2 and RS3 are applied to another data set (DS2) which contains up to 800 wind turbines with different relationships (from p1 to p5). This dataset is used to evaluate ZodiacEdge in more complex cases, \ie more complex rule sets. The rules for these sets are presented in the appendices \ref{ruleset1}, \ref{ruleset2} and \ref{ruleset3} and they can be downloaded from our Github paper companion . 
Table \ref{tab:ruleSets} summarises the main characteristics of each rule and data sets.

\begin{table*}
\begin{center}
\begin{tabular}{ c | c c c c c c c c}
  & DS & \#hn & \#rules & \#predicates & EDB Negation & IDB Negation & Aggregation & \#facts\\ 
  \hline
 RS1 & DS1 &5 & 6 & 6  & & & \checkmark &  up to 160.800 \\
 RS2 & DS2 & 14 & 18 & 15 & \checkmark&  & & 424.300 \\
 RS3 & DS2 & 14 & 18 & 15 & &\checkmark & & 423.101 \\
\end{tabular}
\end{center}
\caption{Data and rule set summaries (DS stands for Data set, \#hn= number of hyper-nodes)}
\label{tab:ruleSets}
\end{table*}

\subsection{Results}

\subsubsection{Size overhead}
\begin{figure}[ht]
\centering
\includegraphics[scale=0.42]{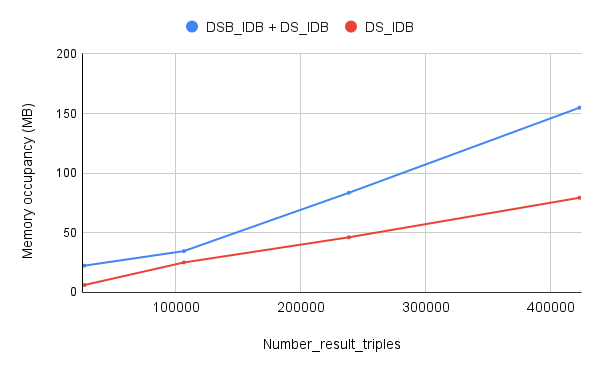}
\caption{Main-memory footprint of DSB and DS data structures}
\label{fig:exp_memory_occupancy}
\end{figure}

In this first experimentation, we evaluate the size overhead of our data structures, \ie DS and DSB, which are mainly used to speed up reasoning. The evaluation is performed on different data sets for RS2, \ie including our most complex use case which contains 14 hyper-nodes. The data sets range from 27K to over 420K facts. We evaluate the ZodiacEdge IDB on memory overhead.
In Fig. \ref{fig:exp_memory_occupancy} we can also see that both data structures grow linearly with the number of hyper-nodes.

\subsubsection{Inference processing}
\begin{figure}[ht]
\centering
\includegraphics[scale=0.4]{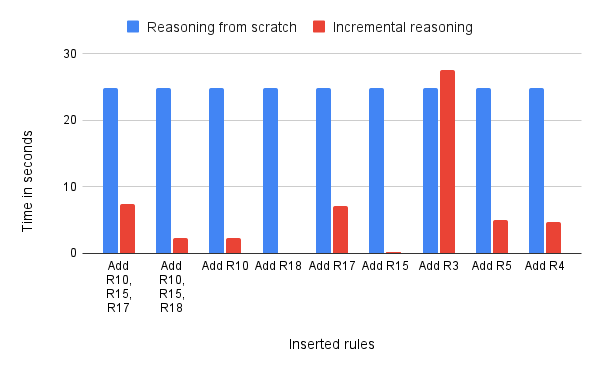}
\caption{Rule-insertion performance with complicated cases.}
\label{fig:exp_rule_incremental_insert}
\end{figure}

\begin{figure}[ht]
\centering
\includegraphics[scale=0.42]{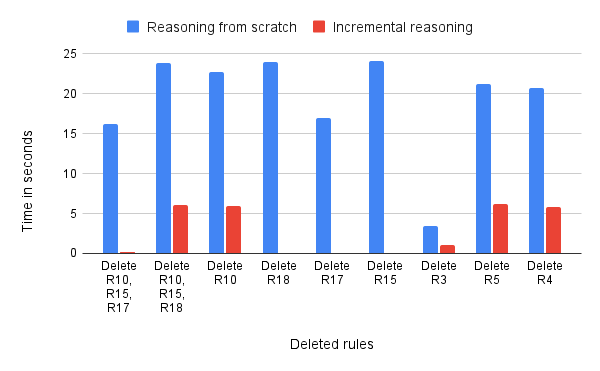}
\caption{Rule-deletion performance with complicated cases.}
\label{fig:exp_rule_incremental_del}
\end{figure}

The goal of this set of experiments is to evaluate the efficiency of our incremental materialization maintenance compared to a naive recalculation of inferences. We are interested in the performance of both insert and delete operations. Our method of experimentation consists of evaluating different sets of more or less complex rules and initializing each experiment with RS2 rules regardless of the rules present in the x-axis of Fig. \ref{fig:exp_rule_incremental_insert} and Fig.\ref{fig:exp_rule_incremental_del} for insertions and deletions respectively. For instance, the first experimentation in Fig. \ref{fig:exp_rule_incremental_insert} initializes 15 of the 18 rules of RS2, \ie excluding r10, r15 and r17, and then adding the latter three rules. The same applies to the second experiment in the same figure, where the three omitted rules are r10, r15 and r18, and then they are added to the rule set. The same approach is applied to the deletion experiment.


Fig. \ref{fig:exp_rule_incremental_insert} highlights that our incremental approach performs better, except for the insertion of r3, than a complete reprocessing of inferences by 3x to 375x depending on the case. The most favorable case is obviously the insertion of r18 which is a leaf hyper-node in RS2's HRDG. Hence, it only needs to point to the EDB of r15 and r16 which is almost instantaneous. 
This confirms that the closer our processing is to a HRDG sheet, the more efficient our approach is. Nevertheless, even in the case of the insertion of r4, our approach is quite efficient with a 5x improvement over a complete new reasoning. The only negative result we get from this experiment is for the insertion of r3, which is recursive and depends on r2. The insertion of r3 creates a hyper-node that involves symmetric-transitive reasoning, which is the most computationally intensive case of reasoning we can get in the language we support. Doing incremental maintenance on such a hyper-node is quite expensive, and may even cost more than redoing the reasoning.



Fig.\ref{fig:exp_rule_incremental_del} highlights that our approach is even more favourable for deletion operations. We get an improvement of 3x to 3 orders of magnitude (for HRDG leaf hyper-nodes).

As for rule deletion, we first initialize ZodiacEdge with the set of rules, after which we delete the rules indicated in the x-axis to obtain the performance of rule deletion. Next, we compare the performance of rule deletion with the inference reprocessing, \ie reasoning with all rules apart from the rules shown in the x-axis. As we can see in Fig. \ref{fig:exp_rule_incremental_del}, in all cases, the rule-incremental strategy performs better than inference reprocessing.

In order to test ZodiacEdge on negation performance, we replace $r10$ in RS2 with $r10_{new}$, which gives us RS3, the corresponding HRDG is shown in Fig. \ref{fig:RS3_HRDG}. We can see that there exists negative dependencies among certain hyper-nodes. We evaluate this new rule set by inserting and deleting r6 and r10 which can typically impact the reasoning with negation. The result of the insertion is shown in Table \ref{tab:exp_rule_incremental_insert_another}.

\begin{table}
\begin{center}
\begin{tabular}{ c | c c }
  & Reasoning from scratch & Incremental \\ 
  \hline
 Add r6 & 24,134 s & 0,2789 s \\  
 Add $r10_{new}$ & 24,134 s & 2,335 s
\end{tabular}
\end{center}
\caption{Insertion performance on Rule Set 3}
\label{tab:exp_rule_incremental_insert_another}
\end{table}

\begin{table}
\begin{center}
\begin{tabular}{ c | c c }
  & Reasoning from scratch & Incremental \\ 
  \hline
 Delete r6 & 24,7 s & 7,304 s \\  
 Delete $r10_{new}$ & 23,347 s & 6,073 s
\end{tabular}
\end{center}
\caption{Deletion performance on Rule Set 3}
\label{tab:exp_rule_incremental_delete_another}
\end{table}
From the previous evaluation, we found that the hyper-node with symmetric-transitive features is always the most computationally intensive part during the reasoning process.

\begin{figure}[ht]
\centering
\includegraphics[scale=0.38]{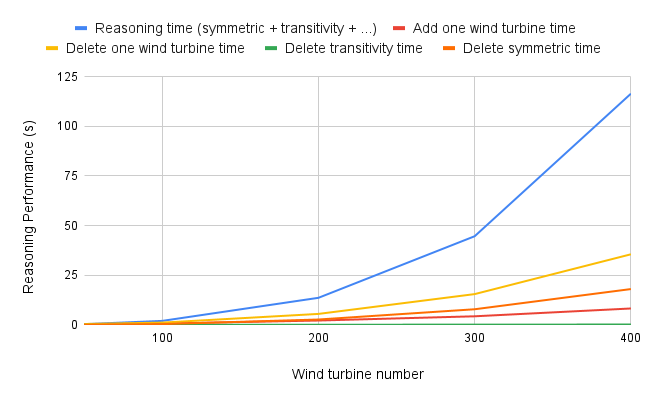}
\caption{Reasoning performance with different operations.}
\label{fig:exp_Reasoning_curve}
\end{figure}

Moreover, it consumes much more computer resources than other forms of reasoning. This motivates the following evaluation of DS1 and RS1, where symmetric-transitive reasoning plays the most important role. We first perform the full reasoning process (symmetric + transitive), and then evaluate the performance of ZodiacEdge in the incremental aspects of the data by adding and removing a triple. After that, we also evaluate the rule deletion performance by deleting the symmetric rule or the transitive rule. Fig. \ref{fig:exp_Reasoning_curve} shows the result. The figure shows that the performance of symmetric-transitive reasoning is proportional to $n^2$, where $n$ represents the number of wind turbines in the data set. This makes sense because the number of relations of a symmetric-transitive property in a fully connected graph is proportional to the square of the number of nodes.

\begin{figure}[ht]
\centering
\includegraphics[scale=0.4]{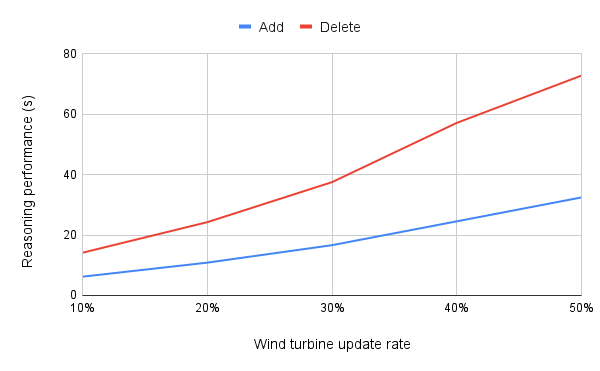}
\caption{Data-incremental Reasoning performance.}
\label{fig:exp_data_incre_reasoning_curve}
\end{figure}

We also perform another experiment on the data-incremental phase. Our data-incremental evaluation is based on a 200 wind turbines data set with symmetric-transitive reasoning requirements. We evaluate ZodiacEdge's performance by first inserting from $10\%$ to $50\%$ extra wind turbines and then deleting these inserted wind turbines. The result of this evaluation can be found in Fig. \ref{fig:exp_data_incre_reasoning_curve}, we can see that the deletion takes more time than the insertion. This is natural because during incremental deletion, ZodiacEdge consumes extra time to check if the deleting triples are no more supported by the reduced EDB. This result was confirmed by the B/F paper.

\section{Conclusion}
\label{conclusion}
In this paper, we presented, to the best of our knowledge, the first work that addresses the incremental maintenance when rules are inserted into or deleted from a Datalog program. These rules can contain negated EDB and IDB predicates, data binding, aggregation as well as comparison operations, such as equality and difference, between predicate terms. Our approach, based on stratification and a dependency hypergraph, has been evaluated on different use cases and presents encouraging results which outperformed a complete inference process by up to three orders of magnitude.

As future work, we would like to adopt an approach similar to \cite{DBLP:journals/ai/HuMH22} where certain rule patterns are addressed with different strategies. We believe that we can further adapt ZodiacEdge's behaviour to certain combinations of rules and that this will be a source of performance improvement.
Another line of work consists of proposing an execution of the rules optimised by dynamic programming and integrating a bookkeeping approach into the incremental maintenance of the materialization.

\bibliographystyle{ACM-Reference-Format}
\bibliography{main}

\appendix
\section{Rule set 1}
\label{ruleset1}
This rule set corresponds to an extract of the real-world Wind farm use case. Due to confidentiality reasons, we can not provide the complete rule set.

r1: hasNeighbour(X, Y) :- hasNeighbour(Y, X) .\\
r2: hasNeighbour(X, Y) :- hasNeighbour(X, Z) $\wedge$ hasNeighbour(Z, Y) $\wedge$ COMP(X, $!$=, Y) .\\
r3: hasNeighbourAirTemperatureMeasurementNumber(X, Z) :- aggregate( hasNeighbour(X, Y) $\wedge$ hasAirTemperatureMesurement(Y, T)) on X with count(T) as Z .\\
r4: hasMedianAirTemperatureMeasurementNearby(X, Z) :- aggregate( hasNeighbour(X, Y) $\wedge$ hasAirTemperatureMesurement(Y, T)) on X with Med(T) as Z .\\
r5: MoreThan3Neighbours(X) :- hasNeighbourAirTemperatureMeasurementNumber(X, N) $\wedge$ Comp(N, >=, 3) .\\
r6: SensorAnomalyWindTurbine(X) :- hasMedianAirTemperatureMeasurementNearby(X, M) $\wedge$ MoreThan3Neighbours(X) $\wedge$ hasAirTemperatureMesurement(X, T) and bind(abs(T-M) as D) $\wedge$ Comp(D,>,5).\\

\begin{figure*}[h]
\centering
\includegraphics[scale=0.30]{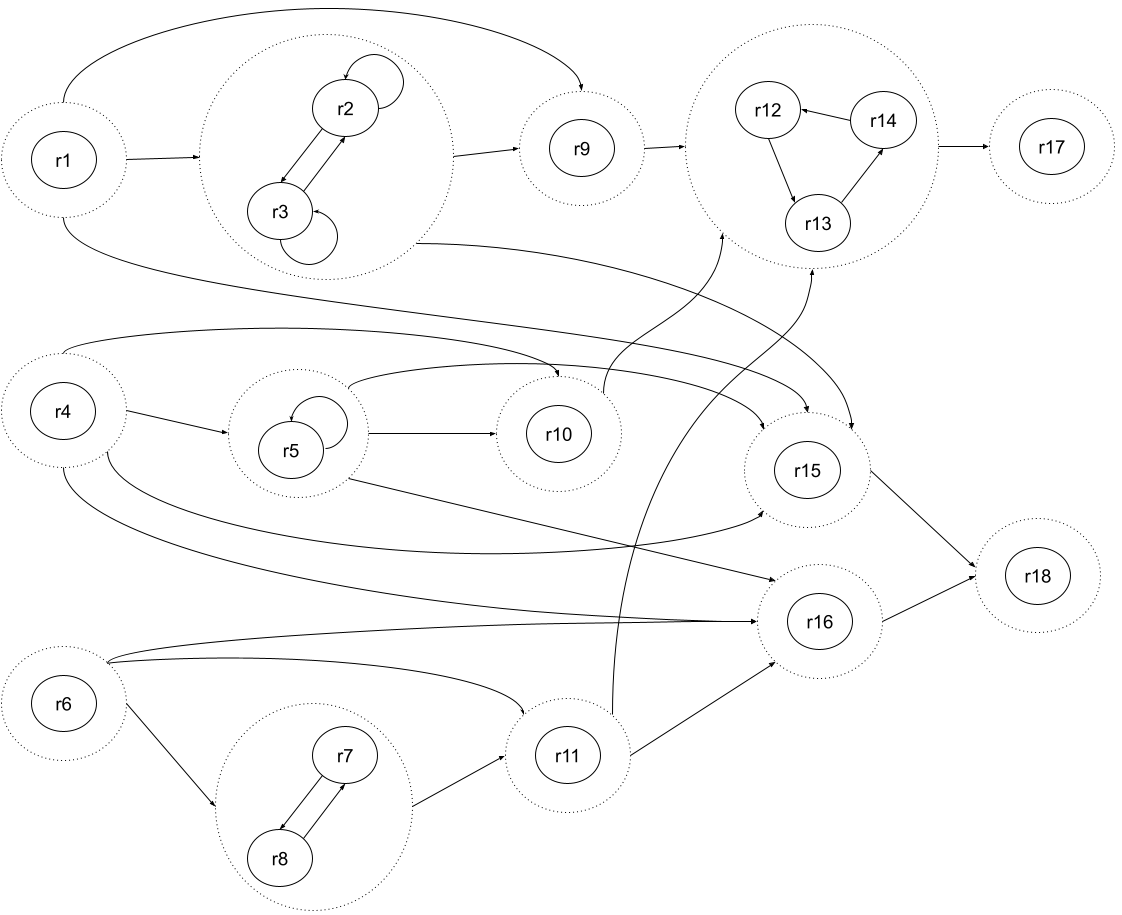}
\caption{Rule set 2's HRDG}
\label{fig:RS2_HRDG}
\end{figure*}

\section{Rule set 2}
\label{ruleset2}

This rule set corresponds to a synthetic use case and is used in the Evaluation section of the paper. It corresponds to a semipositive datalog with negation. It contains symmetric, transitive rules as well as rules with atom inequality and negation on an EDB predicate. Fig. \ref{fig:RS2_HRDG} presents the HRDG of this rule set.\\
r1:	p11(X, Y) :- p1(X, Y).\\		
r2:	p11(X, Y) :- p11(Y, X).\\		
r3:	p11(X, Y) :- p11(X, Z) $\wedge$ p11(Z, Y) $\wedge$ COMP(X,!=, Y).\\		
r4:	p12(X, Y) :- p2(X, Y).\\		
r5:	p12(X, Y) :- p12(X, Z) $\wedge$ p12(Z, Y) 
$\wedge$ COMP(X,!=, Y).\\		
r6:	p13(X, Y) :- p3(X, Y).\\		
r7:	p14(X, Y) :- p13(X, Y).\\		
r8:	p13(X, Y) :- p14(Y, X).\\		
r9:	p20(X, Y) :- p11(X, Y).\\		
r10: p20(X, Y) :- p12(X, Y).\\		
r11: p20(X, Y) :- p13(X, Y).\\		
r12: p21(X, Y) :- p20(X, Y).\\		
r13: p22(X, Y) :- p21(X, Y).\\		
r14: p20(X, Y) :- p22(X, Y).\\		
r15: p25(X, Z) :- p11(X, Y) $\wedge$ p12(Y, Z) $\wedge$ not p5(Y, Z).\\	
r16: p26(X, Z) :- p12(X, Y) $\wedge$ p13(Z, Y) $\wedge$ not p5(Z, Y).\\
r17: p30(X, Z) :- p22(X, Y) $\wedge$ p21(Y, Z).\\
r18: p31(X, Y) :- p25(X, Y) $\wedge$ p26(Y, Z).\\

\begin{figure*}[ht]
\centering
\includegraphics[scale=0.3]{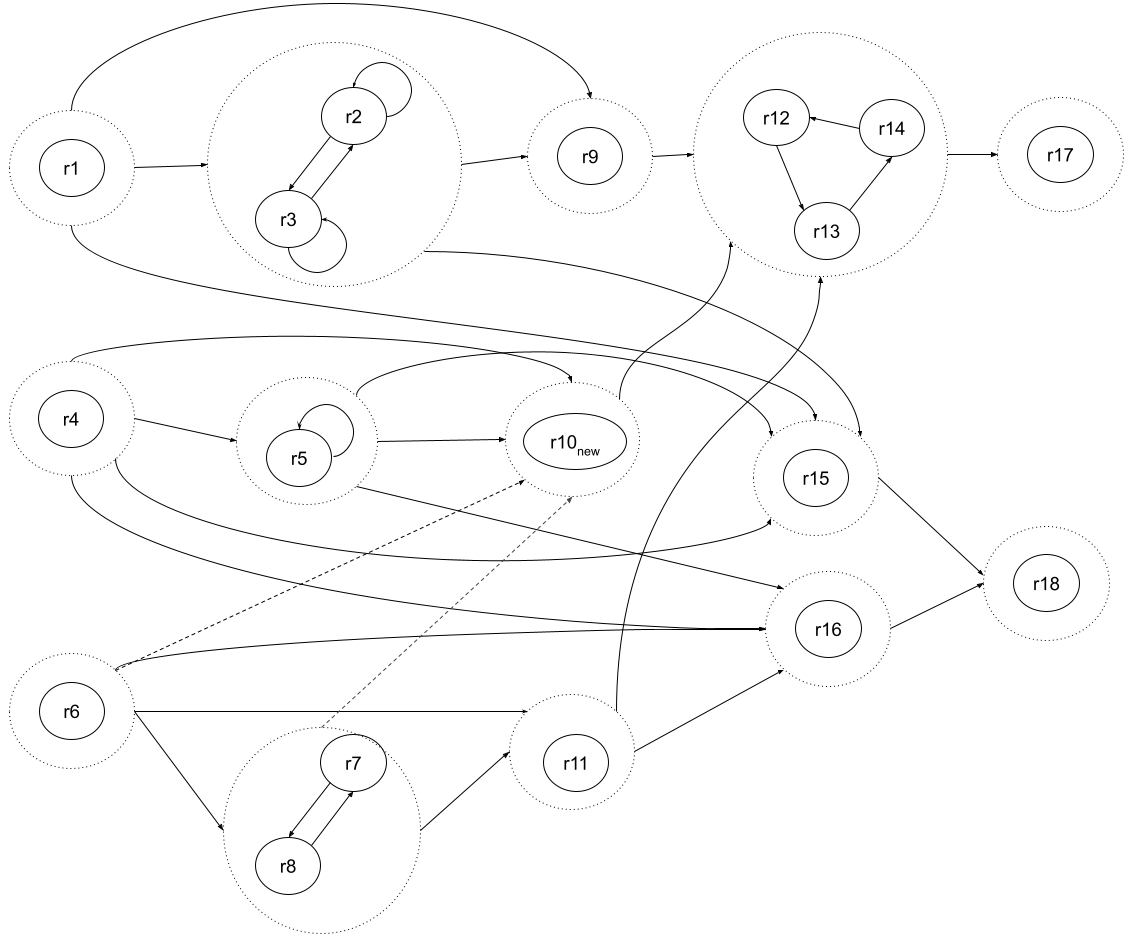}
\caption{Rule set 3's HRDG}
\label{fig:RS3_HRDG}
\end{figure*}

\section{Rule set 3}
\label{ruleset3}

Rule Set 3 is an adaptation of Rule Set 2 with only one modification on rule r10. The Datalog program thus accepts negation on IDB predicates and is not a semipositive datalog program anymore. Fig. \ref{fig:RS3_HRDG} presents this rule set's HRDG.\\
r1:	p11(X, Y) :- p1(X, Y).\\		
r2:	p11(X, Y) :- p11(Y, X).\\		
r3:	p11(X, Y) :- p11(X, Z) $\wedge$ p11(Z, Y) $\wedge$ COMP(X,!=, Y).\\		
r4:	p12(X, Y) :- p2(X, Y).\\		
r5:	p12(X, Y) :- p12(X, Z) $\wedge$ p12(Z, Y) 
$\wedge$ COMP(X,!=, Y).\\		
r6:	p13(X, Y) :- p3(X, Y).\\		
r7:	p14(X, Y) :- p13(X, Y).\\		
r8:	p13(X, Y) :- p14(Y, X).\\		
r9:	p20(X, Y) :- p11(X, Y).\\		
$r10_{new}$: p20(X, Y) :- p12(X, Y) $\wedge$ not p13(Y, Z).\\
r11: p20(X, Y) :- p13(X, Y).\\		
r12: p21(X, Y) :- p20(X, Y).\\		
r13: p22(X, Y) :- p21(X, Y).\\		
r14: p20(X, Y) :- p22(X, Y).\\		
r15: p25(X, Z) :- p11(X, Y) $\wedge$ p12(Y, Z) $\wedge$ not p5(Y, Z).\\	
r16: p26(X, Z) :- p12(X, Y) $\wedge$ p13(Z, Y) $\wedge$ not p5(Z, Y).\\
r17: p30(X, Z) :- p22(X, Y) $\wedge$ p21(Y, Z).\\
r18: p31(X, Y) :- p25(X, Y) $\wedge$ p26(Y, Z).\\

\end{document}